\documentclass[notitlepage]{article}
\usepackage{amsmath}
\usepackage{amsfonts}
\usepackage{amssymb}
\newtheorem{theorem}{Theorem}

\newtheorem{corollary}[theorem]{Corollary}

\newtheorem{definition}[theorem]{Definition}

\newtheorem{notation}[theorem]{Notation}

\newtheorem{remark}[theorem]{Remark}

\newenvironment{proof}[1][Proof]{\textbf{#1.} }{\ \rule{0.5em}{0.5em}}

\begin{document}

\title{Operator-Schmidt decomposition of the \\quantum Fourier transform on $\mathbb{C}^{N_{1}}\otimes\mathbb{C}^{N_{2}}$}
\author{Jon Tyson\thanks{Email: jonetyson@post.harvard.edu}\\Jefferson Lab, Harvard University, Cambridge, MA 02138 USA\footnote{current address}}
\date{April 25, 2003}
\maketitle
\begin{abstract}
\noindent Operator-Schmidt decompositions of the quantum Fourier transform on
$\mathbb{C}^{N_{1}}\otimes\mathbb{C}^{N_{2}}$ are computed for all $N_{1}
$,$N_{2}\geq2$. The decomposition is shown to be completely degenerate when
$N_{1}$ is a factor of $N_{2}$ and when $N_{1}>N_{2}$. The first known special
case, $N_{1}=N_{2}=2^{n},$ was computed by Nielsen in his study of the
communication cost of computing the quantum Fourier transform of a collection
of qubits equally distributed between two parties. [M. A. Nielsen, PhD Thesis,
University of New Mexico (1998), Chapter 6, arXiv:quant-ph/0011036.] More
generally, the special case $N_{1}=2^{n_{1}}\leq2^{n_{2}}=N_{2}$ was computed
by Nielsen et. al. in their study of strength measures of quantum operations.
[M. A. Nielsen et. al, (accepted for publication in \textit{Phys. Rev. A});
arXiv:quant-ph/0208077.] Given the Schmidt decompositions presented here, it
follows that in all cases the bipartite communication cost of exact
computation of the quantum Fourier transform is maximal.

\medskip\noindent PACS numbers:\ 03.67.-a
\end{abstract}

\pagebreak 

\section{\label{introsec}Introduction}

\noindent Operator-Schmidt decompositions are useful for quantifying the
non-local nature of operators on finite-dimensional bipartite Hilbert spaces.
The first special cases of Schmidt decompositions of the quantum Fourier
transform were computed by Nielsen $\cite{NielsenThesis}$ to illustrate his
study of coherent quantum communication complexity. He considered the
following problem:\smallskip

\begin{quotation}
\noindent Suppose Alice is in possession of $m$ qubits, Bob is in possession
of $n$ qubits, and they wish to perform some general unitary operation $U$
which acts on their $m+n$ qubits. How many qubits must be communicated between
Alice and Bob for them to achieve this goal?
\end{quotation}

\noindent Nielsen proved that the number $Q_{0}\left(  U\right)  $ of such
qubits was bounded by
\begin{equation}
1/2\times K_{\text{Har}}\left(  U\right)  \leq Q_{0}\left(  U\right)
\leq2\min\left(  n,m\right)  ,\label{weakboundnielsen}%
\end{equation}
where the \textit{Hartley strength }$K_{\text{Har }}$ satisfies
\[
K_{\text{Har}}\left(  U\right)  \equiv\log_{2}\left(  \operatorname{Sch}%
\left(  U\right)  \right)  \text{,}%
\]
where $\operatorname{Sch}\left(  U\right)  ,$ defined in Definition
\ref{schmidtdef} below, is the number of nonzero Schmidt coefficients of $U$.
The upper bound of $\left(  \ref{weakboundnielsen}\right)  $ is trivial, for
Alice could simply send her qubits to Bob and let him send them back after
performing $U, $ or vice-versa. To illustrate his theorem, Nielsen considered
the quantum Fourier transform $\mathcal{F}_{2^{n}\times2^{n}}$ on $n+n$
qubits. He showed that $K_{\text{Har}}\left(  \mathcal{F}_{2^{n}\times2^{n}%
}\right)  =2n, $ yielding $n\leq Q_{0}\left(  \mathcal{F}_{2^{n}\times2^{n}%
}\right)  \leq2n$. Subsequent work by Nielsen \cite{nielsenemail} improved the
\textit{general} lower bound of $\left(  \ref{weakboundnielsen}\right)  $ by a
factor of two,\footnote{See also footnote \ref{alternativefootnote} for a
brief outline of an alternative proof. We remark that Nielsen considers qubits
for convenience only. In particular, let $V$ be a unitary on $\mathbb{C}%
^{N_{1}}\otimes C^{N_{2}}$, where $N_{1}$ and $N_{2}$ are the respective
dimensions of Alice and Bob's quantum states, with no requirement that $N_{1}$
and $N_{2}$ be a powers of two. Then any quantum computation of $V$ employing
some combination of qudit communication and ancillae, possibly of varying
dimension, satisfies the following bound: $\sum_{d=2}^{\infty}N_{d}\log
_{2}\left(  d\right)  \geq K_{\text{har}}\left(  V\right)  $, where $N_{d}$ is
the number of qudits of dimension $d$ communicated between Alice and Bob. It
is assumed that at the end of the computation that Alice and Bob retain
posession of their (now altered) data qudits, although the bound holds whether
or not a given net transfer of the (restored) ancillae is allowed.} in
particular implying that
\[
Q_{0}\left(  \mathcal{F}_{2^{n}\times2^{n}}\right)  =2n.
\]

In a later paper \cite{DynamicalStrength}, Nielsen and collaborators further
employ operator Schmidt decompositions in the quantitative study of
\textit{strength measures} of the nonlocal action of unitary
operators.\footnote{They also consider more general quantum operations than
unitaries.} Besides revisiting the Hartley strength, among the several
strength measures considered is the \textit{Schmidt strength},
\[
K_{\text{Sch}}\left(  U\right)  =H\left(  \left\{  \frac{\lambda_{k}^{2}}%
{\dim\left(  \mathcal{H}\otimes\mathcal{K}\right)  }\right\}  \right)
\text{,}%
\]
where $U$ is a unitary operator on $\mathcal{H}\otimes\mathcal{K}$, $\left\{
\lambda_{k}\right\}  $ are its Schmidt coefficients, and $H$ is the Shannon
entropy. They give a Schmidt decomposition of $\mathcal{F}_{2^{m}\times2^{n}}$
on $m+n$ qubits for the case $m\leq n$ and conjecture that $\operatorname{Sch}%
\left(  \mathcal{F}_{2^{m}\times2^{n}}\right)  =2^{2n}$ for $m>n.$

\subsection{Results}

Schmidt decompositions of the quantum Fourier transform $\mathcal{F}%
_{N_{1}\times N_{2}}:\mathbb{C}^{N_{1}}\otimes\mathbb{C}^{N_{2}}%
\rightarrow\mathbb{C}^{N_{1}}\otimes\mathbb{C}^{N_{2}}$ are given for all
$N_{1},N_{2}>1$, with no requirement that either $N_{1}$ or $N_{2}$ be a power
of two. As a special case, the conjecture of Nielsen and collaborators is
affirmed. In all cases, the results of Nielsen imply that the bipartite
communication cost of exact computation of the quantum Fourier transform is
maximal. Once stated, the decomposition is easily verified; a short derivation
is given in the Appendix.\smallskip

\subsection{\label{natiso}Definitions and Notation}

\begin{definition}
Let $N,$ $N_{1},$ $N_{2}$ be integers greater than one satisfying
$N=N_{1}N_{2}$. The \textbf{quantum Fourier transformation}\footnote{This is
unitarily equivalent to the discrete Fourier transform.} $\mathcal{F}%
_{N}:\mathbb{C}^{N}\rightarrow\mathbb{C}^{N}$ is the unitary operator
satisfying
\[
\mathcal{F}_{N}\;\left|  s\right\rangle _{N}=\frac{1}{\sqrt{N}}\sum
_{t=0}^{N-1}\exp\left(  \frac{2\pi i}{N}ts\right)  \left|  t\right\rangle
_{N},\;s\in\left\{  0,...,N-1\right\}  ,
\]
where each $\left|  s\right\rangle _{N}$ is a standard basis element. The
quantum Fourier transformation $\mathcal{F}_{N_{1}\times N_{2}}$ on
$\mathbb{C}^{N_{1}}\otimes\mathbb{C}^{N_{2}}$ is obtained by identifying
$\mathbb{C}^{N}$ with $\mathbb{C}^{N_{1}}\otimes\mathbb{C}^{N_{2}}$ under the
\textbf{mixed-decimal representation}, which asserts the equalities
\[
\left|  s\right\rangle _{N}=\left|  k\ell\right\rangle _{N_{1},N_{2}}=\left|
k\right\rangle _{N_{1}}\otimes\left|  \ell\right\rangle _{N_{2}}%
\]
when
\[
s=kN_{2}+\ell\text{, }k\in\left\{  0,...,N_{1}-1\right\}  \text{, }\ell
\in\left\{  0,...,N_{2}-1\right\}  \text{.}%
\]
\end{definition}

\begin{remark}
In the case that $N_{1}\neq N_{2}$, the reader is warned that the operator
$\mathcal{F}_{N_{1}\times N_{2}}$ is not equivalent in what follows to
$\mathcal{F}_{N_{2}\times N_{1}}$. Specifically, $\mathcal{F}_{N}$ does not
commute with the unitary operator $R_{N_{1},N_{2}}:\mathbb{C}^{N}%
\rightarrow\mathbb{C}^{N}$ given by
\[
R_{N_{1},N_{2}}\left|  kN_{2}+\ell\right\rangle =\left|  \ell N_{1}%
+k\right\rangle \text{,\ \ \ }k\in\left\{  0,...,N_{1}-1\right\}  \text{,
}\ell\in\left\{  0,...,N_{2}-1\right\}  ,
\]
which interchanges the digits of the mixed-decimal representation.
\end{remark}

\begin{notation}
Let $\mathcal{H}$ be a finite-dimensional Hilbert space. Then $B\left(
\mathcal{H}\right)  $ is the Hilbert space of linear transformations on
$\mathcal{H}$ with the Hilbert-Schmidt inner product $\left\langle
A,B\right\rangle _{B\left(  \mathcal{H}\right)  }=\operatorname*{Tr}A^{%
\dag
}B$.\footnote{If $A$ is a linear operator on $\mathcal{H}$, then $A^{%
\dag
}$ is defined by $\left\langle x,Ay\right\rangle _{\mathcal{H}}=\left\langle
A^{%
\dag
}x,y\right\rangle _{\mathcal{H}}$ for all $x,y\in\mathcal{H}$. Here
$\left\langle \bullet,\bullet\right\rangle _{\mathcal{H}}$ is the inner
product on $\mathcal{H}$, and we will always take inner products to be linear
in the second argument.}
\end{notation}

\begin{definition}
\label{schmidtdef}Let $\mathcal{H}$ and $\mathcal{K}$ be finite-dimensional
Hilbert spaces, and let $F$ be a nonzero linear transformation on
$\mathcal{H}\otimes\mathcal{K}$. An \textbf{(operator) Schmidt decomposition}
of $F$ is a decomposition of the form
\begin{equation}
F=\sum_{k=1}^{\operatorname{Sch}\left(  F\right)  }\lambda_{k}\;A_{k}\otimes
B_{k}\text{,\ \ }\lambda_{k}>0,\text{\label{opschmform}}%
\end{equation}
where $\left\{  A_{k}\right\}  _{k=1...\operatorname{Sch}\left(  F\right)  }$
and $\left\{  B_{k}\right\}  _{k=1...\operatorname{Sch}\left(  F\right)  }$
are orthonormal sets\footnote{but not necessarily bases} of operators on
$\mathcal{H}$ and $\mathcal{K}$ respectively, under the Hilbert-Schmidt inner
product. The quantity\textbf{\ }$\operatorname{Sch}\left(  F\right)  $ is
called the \textbf{Schmidt number, }and the $\lambda_{k}$ are called the
\textbf{Schmidt coefficients}. Such a decomposition is said to be
\textbf{completely degenerate }if $\operatorname{Sch}\left(  F\right)
=\left(  \min\left(  \dim\mathcal{H},\dim\mathcal{K}\right)  \right)  ^{2}$
and all the $\lambda_{k}$ are equal.\footnote{\label{alternativefootnote}More
generally, if $F:\mathcal{H}\otimes\mathcal{K}\rightarrow\mathcal{H}^{\prime
}\otimes\mathcal{K}^{\prime}$, then one may consider decompositions of the
form $\left(  \ref{opschmform}\right)  $, where now the $A_{k}:\mathcal{H}%
\rightarrow\mathcal{H}^{\prime}$ and $B_{k}:\mathcal{K}\rightarrow
\mathcal{K}^{\prime}$ are orthornormal. A useful such decomposition exists for
the \textit{communication operator} $C:\left(  \mathbb{C}^{n_{1}}%
\otimes\mathbb{C}^{n_{2}}\right)  \otimes\mathbb{C}^{n_{3}}\rightarrow
\mathbb{C}^{n_{1}}\otimes\left(  \mathbb{C}^{n_{2}}\otimes\mathbb{C}^{n_{3}%
}\right)  $, defined by $C\left(  a\otimes b\right)  \otimes c=a\otimes\left(
b\otimes c\right)  $. One may check that $C=\sum_{k=1}^{n_{2}}\sqrt{n_{1}%
n_{3}}A_{k}\otimes B_{k}$, where $A_{k}=n_{1}^{-1/2}\sum_{i=1}^{n_{1}}\left|
i\right\rangle \left\langle ik\right|  :$ $\mathbb{C}^{n_{1}}\otimes
\mathbb{C}^{n_{2}}\rightarrow\mathbb{C}^{n_{1}}$ and $B_{k}=n_{3}^{-1/2}%
\sum_{i=1}^{n_{3}}\left|  ki\right\rangle \left\langle i\right|
:\mathbb{C}^{n_{3}}\rightarrow\mathbb{C}^{n_{2}}\otimes\mathbb{C}^{n_{3}}$.
Replacing the swap operators in section III.B.3 of \cite{DynamicalStrength} by
communication operators, one obtains the aforementioned sharp quantum
communication complexity bound of \cite{nielsenemail}.}
\end{definition}

We remark that the operator-Schmidt decomposition is just a special case of
the well-known Schmidt-decomposition
\[
\psi=\sum_{k=1}^{\operatorname{Sch}\left(  \psi\right)  }\lambda_{k}%
\,e_{k}\otimes f_{k}\text{, \ }\lambda_{k}>0
\]
of a vector $\psi\in\mathcal{H}_{0}\otimes\mathcal{K}_{0}$, where the
$\left\{  e_{k}\right\}  $ and $\left\{  f_{k}\right\}  $ are
orthonormal.\footnote{See $\cite{nielsenbook}$ for a discussion of the Schmidt
decomposition.}\smallskip\ In particular, one sets $\mathcal{H}_{0}=B\left(
\mathcal{H}\right)  ,$ $\mathcal{K}_{0}=B\left(  \mathcal{K}\right)  $, and
$\psi=F\in B\left(  \mathcal{H}\otimes\mathcal{K}\right)  $. The decomposition
$\left(  \ref{opschmform}\right)  $ is then obtained by identifying $B\left(
\mathcal{H}\right)  \otimes B\left(  \mathcal{K}\right)  $ with $B\left(
\mathcal{H}\otimes\mathcal{K}\right)  $ under the natural
isomorphism.\footnote{In particular, there exists a unique unitary
$\Xi:B\left(  \mathcal{H}\right)  \otimes B\left(  \mathcal{K}\right)
\rightarrow B\left(  \mathcal{H}\otimes\mathcal{K}\right)  $ such that
$\left(  \Xi\left(  A\tilde{\otimes}B\right)  \right)  \left(  f\otimes
g\right)  =\left(  Af\right)  \otimes\left(  Bg\right)  $ for all
$f\in\mathcal{H}$ and $g\in\mathcal{K}$. Here $\tilde{\otimes}$ denotes the
defining tensor product of $B\left(  \mathcal{H}\right)  \otimes B\left(
\mathcal{K}\right)  $, considering $B\left(  \mathcal{H}\right)  $ and
$B\left(  \mathcal{K}\right)  $ as abstract Hilbert spaces.} It follows that
$F$ and $G$ in $B\left(  \mathcal{H}\otimes\mathcal{K}\right)  \simeq$
$B\left(  \mathcal{H}\right)  \otimes B\left(  \mathcal{K}\right)  $ have the
same operator-Schmidt coefficients, counting multiplicity, iff
\[
A=\left(  \mathbb{U}\otimes\mathbb{V}\right)  B
\]
for some unitary ``super-operators''$\;\mathbb{U}\in B\left(  B\left(
\mathcal{H}\right)  \right)  $ and $\mathbb{V}\in B\left(  B\left(
\mathcal{K}\right)  \right)  $.\footnote{See exercise 2.80 of
\cite{nielsenbook}. One would like to know much more, i.e. invariants which
specify when are there local unitaries $U,Y\in B\left(  \mathcal{H}\right)  $
and $V,Z\in B\left(  \mathcal{K}\right)  $ such that $A=\left(  U\otimes
V\right)  B\left(  Y\otimes Z\right)  $. Such invariants are known only in the
two-qubit case, \cite{makhlin} where one has the corresponding canonical
decomposition of Khaneja, Brockett, and Glaser \cite{firstcanonicaldecomp}
(see also Kraus and Cirac \cite{krauscirac} for simple ``magic basis'' proof.)}

The well-known procedure for computing Schmidt decompositions is reviewed in
Theorem \ref{schmidtsummarytheo} of the Appendix.\ We content ourselves here
with the statement that the Schmidt coefficients of $\psi\in\mathcal{H}%
_{0}\otimes\mathcal{K}_{0}$ are the square roots of the nonzero eigenvalues of
the reduced density matrix
\[
\rho_{\psi}=\operatorname*{Tr}_{\mathcal{K}_{0}}\left|  \psi\right\rangle
\left\langle \psi\right|  \text{.}%
\]
Equivalently, the Schmidt coefficients are the nonzero singular values of the
operator $B_{\psi}:\mathcal{H}_{0}\rightarrow\mathcal{K}_{0}^{\ast}$ given by
\[
\left(  B_{\psi}f\right)  \left(  g\right)  =\left\langle \psi,f\otimes
g\right\rangle _{\mathcal{H}_{0}\otimes\mathcal{K}_{0}},
\]
where $\mathcal{K}_{0}^{\ast}$ is the dual space of continuous linear
functionals on $\mathcal{K}_{0}$.\footnote{See \cite{nielsenbook} for a proof
that the Schmidt decomposition is a consequence of the singular value
decomposition. In fact they are mathematically equivalent.}

\section{Schmidt Decomposition of $\mathcal{F}$}

\begin{notation}
Let $\mathbb{Z}_{N_{1}}=\left\{  0,...,N_{1}-1\right\}  $, $\mathbb{Z}_{N_{2}%
}=\left\{  0,...,N_{2}-1\right\}  $, $\mathbb{Z}_{N_{2}}^{2}=\mathbb{Z}%
_{N_{2}}\times\mathbb{Z}_{N_{2}}$, $N_{1}\mathbb{Z}^{2}=\left\{  \left(
N_{1}x,N_{1}y\right)  \,|\,x,y\in\mathbb{Z}\right\}  $, $\left\lceil
x\right\rceil =\min\left\{  n\in\mathbb{Z}\;|\;n>x\right\}  $, and
$\left\lfloor x\right\rfloor =-\left\lceil -x\right\rceil $. Denote the
cardinality of a set $C$ by $\left|  C\right|  $. Its \textbf{characteristic
function }$\chi_{C}$ satisfies
\[
\chi_{C}\left(  x\right)  =\left\{
\begin{array}
[c]{ccc}%
1 & \text{if} & x\in C\\
0 & \text{if} & x\notin C
\end{array}
\right.  \text{.}%
\]
Adopt the convention $n\,\operatorname{mod}m\in\mathbb{Z}_{m}.$
\end{notation}

\begin{theorem}
\label{anactualbasis}Define an equivalence relation $\sim$ on $\mathbb{Z}%
_{N_{2}}^{2}$ by
\[
\vec{\ell}\sim\vec{m}\Longleftrightarrow\vec{\ell}-\vec{m}\in N_{1}%
\mathbb{Z}^{2}\text{,}%
\]
where the subtraction is \textit{not} modular, and define $\mathcal{M}%
=\mathbb{Z}_{N_{2}}^{2}/\sim$ to be the set of equivalence
classes.\footnote{The reader may check that for $N_{1}=2$ and $N_{2}=3$ that
$\mathcal{M}$ consists of $\left\{  \left(  0,0\right)  ,\left(  0,2\right)
,\left(  2,0\right)  ,\left(  2,2\right)  \right\}  $, $\left\{  \left(
1,0\right)  ,\left(  1,2\right)  \right\}  $, $\left\{  \left(  0,1\right)
,\left(  2,1\right)  \right\}  $, and $\left\{  \left(  1,1\right)  \right\}
$.} Then a Schmidt decomposition of $\mathcal{F}_{N_{1}\times N_{2}}$ is given
by
\begin{equation}
\mathcal{F}_{N_{1}\times N_{2}}=\sum_{C\in\mathcal{M}}\sqrt{\frac{N_{1}}%
{N_{2}}\left|  C\right|  }\;A_{C}\otimes B_{C},\label{thefdecomp}%
\end{equation}
where the matrices of $A_{C}:\mathbb{C}^{N_{1}}\rightarrow\mathbb{C}^{N_{1}} $
and $B_{C}:\mathbb{C}^{N_{2}}\rightarrow\mathbb{C}^{N_{2}}$ are defined by
\[%
\begin{array}
[c]{ll}%
\left(  A_{C}\right)  _{k_{1}k_{2}}=\frac{1}{N_{1}}\exp\left[  \frac{2\pi
i}{N_{1}}\left(  N_{2}k_{1}k_{2}+k_{1}\tilde{c}_{2}+k_{2}\tilde{c}_{1}\right)
\right]  \smallskip &  k_{1},k_{2}\in\mathbb{Z}_{N_{1}}\\
\left(  B_{C}\right)  _{\ell_{1}\ell_{2}}=\frac{1}{\left|  C\right|  ^{1/2}%
}\exp\left(  \frac{2\pi i}{N}\ell_{1}\ell_{2}\right)  \chi_{C}\left(  \left(
\ell_{1},\ell_{2}\right)  \right)  & \ell_{1},\ell_{2}\in\mathbb{Z}_{N_{2}},
\end{array}
\]
with each $\left(  \tilde{c}_{1},\tilde{c}_{2}\right)  \in C$ arbitrarily
chosen. ($A_{C}$ doesn't depend on this choice.)
\end{theorem}

\begin{proof}
It is trivial to check that $\left\{  A_{C}\right\}  $ and $\left\{
B_{C}\right\}  $ are orthonormal sets. Furthermore, for $k_{1},k_{2}%
\in\mathbb{Z}_{N_{1}}$ and $\ell_{1},\ell_{2}\in\mathbb{Z}_{N_{2}}$,%
\begin{align*}
\left\langle k_{1},\ell_{1}\right| & \left(  \sum_{C\in\mathcal{M}}\sqrt
{\frac{N_{1}}{N_{2}}\left|  C\right|  }\,A_{C}\otimes B_{C}\right)
\,\left|  k_{2},\ell_{2}\right\rangle\\
& =\sum_{C\in\mathcal{M}}\sqrt{\frac{N_{1}}{N_{2}}\left|  C\right|
}\,\left\langle k_{1}\right|  A_{C}\left|  k_{2}\right\rangle\left\langle
\ell_{1}\right|  B_{C}\,\left|  \ell_{2}\right\rangle\\
& =\sum_{C\in\mathcal{M}}{\frac{1}{\sqrt{N}}}\exp\left[  \frac{2\pi i}{N}%
\left(
N_{2}^{2}k_{1}k_{2}+N_{2}k_{1}\tilde{c}_{2}+N_{2}k_{2}\tilde{c}_{1}+\ell
_{1}\ell_{2}\right)  \right]  \chi_{C}\left(  \left(  \ell_{1},\ell
_{2}\right)  \right)  \\
& =\sum_{C\in\mathcal{M}}{\frac{1}{\sqrt{N}}}\exp\left[  \frac{2\pi i}{N}%
\left(
N_{2}^{2}k_{1}k_{2}+N_{2}k_{1}\ell_{2}+N_{2}k_{2}\ell_{1}+\ell_{1}\ell
_{2}\right)  \right]  \chi_{C}\left(  \left(  \ell_{1},\ell_{2}\right)
\right)  \\
& ={\frac{1}{\sqrt{N}}}\,\exp\left[  \frac{2\pi i}{N}\left(  N_{2}k_{1}+\ell
_{1}\right)  \left(  N_{2}k_{2}+\ell_{2}\right)  \right]  \\
& =\left\langle k_{1}\ell_{1}\right|  \mathcal{F}_{N_{1}\times N_{2}}\,\left|
k_{2},\ell_{2}\right\rangle,
\end{align*}%
as desired.
\end{proof}

The reader may find it instructive to compute the linear spans of the matrices
$B_{C}$ corresponding to each of the Schmidt coefficients.

\begin{corollary}
\label{putinnumbers}The Schmidt decompositions of $\mathcal{F}_{N_{1}\times
N_{2}}$ fall into three categories:

\begin{enumerate}
\item  If $N_{1}$ is a factor of $N_{2},$ then there is only one Schmidt
coefficient,$\sqrt{N_{2}/N_{1}}$, with multiplicity $N_{1}^{2}$.

\item  If $N_{1}\geq N_{2},$ there is only one Schmidt coefficient,
$\sqrt{N_{1}/N_{2}}$, with multiplicity $N_{2}^{2}$.

\item  Otherwise, $\mathcal{F}_{N_{1}\times N_{2}}$ has three distinct nonzero
Schmidt coefficients:
\[%
\begin{array}
[c]{lll}%
\sqrt{\left\lceil \frac{N_{2}}{N_{1}}\right\rceil ^{2}\frac{N_{1}}{N_{2}}} &
\text{,} & \text{multiplicity }\left(  N_{2}\operatorname{mod}N_{1}\right)
^{2}\\
\sqrt{\left\lceil \frac{N_{2}}{N_{1}}\right\rceil \left\lfloor \frac{N_{2}%
}{N_{1}}\right\rfloor \frac{N_{1}}{N_{2}}} & \text{,} & \text{multiplicity
}2\left(  N_{2}\operatorname{mod}N_{1}\right)  \left(  \left(  -N_{2}\right)
\operatorname{mod}N_{1}\right) \\
\sqrt{\left\lfloor \frac{N_{2}}{N_{1}}\right\rfloor ^{2}\frac{N_{1}}{N_{2}}} &
\text{,} & \text{multiplicity }\left(  \left(  -N_{2}\right)
\operatorname{mod}N_{1}\right)  ^{2}%
\end{array}
\]
\end{enumerate}

\noindent In all cases, the Schmidt number of $\mathcal{F}_{N_{1}\times N_{2}%
}$ is $\min\left(  N_{1}^{2},N_{2}^{2}\right)  $. In particular, the Schmidt
decomposition is completely degenerate in Cases $1$ and $2$.
\end{corollary}

We remark that the previously known cases fall under Case 1. Case 2 verifies
the Schmidt numbers conjectured in \cite{DynamicalStrength}. Since the Schmidt
decomposition in Case 1 (or Case 2) is completely degenerate, Theorem
\ref{schmidtsummarytheo} (below), may be used to find a Schmidt decomposition
of the form of equation $\left(  \ref{opschmform}\right)  $ for \emph{any}
orthonormal basis $\left\{  A_{k}\right\}  $ (or $\left\{  B_{k}\right\}  $,
in case 2).\footnote{Note that cases 1 and 2 overlap for $N_{1}=N_{2}.$}

\ \ \newline \textbf{Acknowledgments:} Michael Nielsen's correspondence was
greatly appreciated. I\ would like to thank Mary Beth Ruskai for her comments,
which were most useful in making the manuscript more readable. This research
was carried out for the Clay Mathematics Institute.

\section{Appendix: A Derivation}

It will soon be apparent that the crucial fact which allows easy calculation
of a Schmidt decomposition of $\mathcal{F}$ is the following:\ \textit{No two
of the }$B_{C}$ \textit{have a nonzero matrix entry in the same place.}

The well-known computational recipe needed here is summarized in

\begin{theorem}
\label{schmidtsummarytheo}Let $\psi\in\mathcal{H}\otimes\mathcal{K}$ be
nonzero. If
\[
\rho_{\mathcal{K}}\equiv\operatorname*{Tr}_{\mathcal{H}}\left|  \psi
\right\rangle \left\langle \psi\right|  =\sum_{\ell\in L}\mu_{\ell}\left|
f_{\ell}\right\rangle \left\langle f_{\ell}\right|  \text{,}%
\]
is a spectral decomposition of the reduced density matrix, then a Schmidt
decomposition of $\psi$ is given by
\begin{equation}
\psi=\sum_{\left\{  \ell|\mu_{\ell}>0\right\}  }\sqrt{\mu_{\ell}}\;e_{\ell
}\otimes f_{\ell}\text{,\label{vschfm}}%
\end{equation}
where each $e_{\ell}$ is defined by the requirement that
\begin{equation}
\left\langle \psi,v\otimes f_{\ell}\right\rangle _{\mathcal{H}\otimes
\mathcal{K}}=\sqrt{\mu_{k}}\left\langle e_{\ell},v\right\rangle _{\mathcal{H}%
}\label{findas}%
\end{equation}
for all $v\in\mathcal{H}$. Furthermore, all Schmidt decompositions of $\psi$
may be exhibited in this manner.
\end{theorem}

\begin{proof}
[Derivation of Theorem \ref{anactualbasis}]We follow the prescription of
Theorem \ref{schmidtsummarytheo}, and employ the natural isomorphism $B\left(
\mathbb{C}^{N_{1}}\right)  \otimes B\left(  \mathbb{C}^{N_{2}}\right)  \simeq
B\left(  \mathbb{C}^{N_{1}}\otimes\mathbb{C}^{N_{2}}\right)  $, as explained
in section \ref{natiso} The reduced density superoperator $\rho\in B\left(
B\left(  \mathbb{C}^{N_{2}}\right)  \right)  $ is defined by the equation
\[
\left\langle A,\rho B\right\rangle _{B\left(  \mathbb{C}^{N_{2}}\right)
}=\sum_{E}\left\langle E\otimes A,\mathcal{F}\right\rangle _{B\left(
\mathbb{C}^{N_{1}}\otimes\mathbb{C}^{N_{2}}\right)  }\left\langle
\mathcal{F},E\otimes B\right\rangle _{B\left(  \mathbb{C}^{N_{1}}%
\otimes\mathbb{C}^{N_{2}}\right)  \text{,}}%
\]
for arbitrary $A,B\in B\left(  \mathbb{C}^{N_{2}}\right)  $, where $E$ runs
over a basis of $B\left(  \mathbb{C}^{N_{1}}\right)  $. For $\vec{j}%
\in\mathbb{Z}_{N_{1}}^{2}$ and $\vec{\ell}\in\mathbb{Z}_{N_{2}}^{2}$define the
standard basis elements
\[
E_{\vec{j}}=\left|  j_{1}\right\rangle \left\langle j_{2}\right|  \in B\left(
C^{N_{1}}\right)  \text{,\ }F_{\vec{\ell}}=\left|  \ell_{1}\right\rangle
\left\langle \ell_{2}\right|  \in B\left(  \mathbb{C}^{N_{2}}\right)  .
\]
We compute $\rho$ by studying its matrix coordinates
\[
\rho_{\vec{\ell}\vec{m}}=\left\langle F_{\vec{\ell}},\rho F_{\vec{m}%
}\right\rangle _{B\left(  \mathbb{C}^{N_{2}}\right)  }\text{,}%
\]
Similarly, let
\[
\mathcal{F}_{\vec{j}\vec{\ell}}=\left\langle E_{\vec{j}}\otimes F_{\vec{\ell}%
},\mathcal{F}\right\rangle _{B\left(  \mathbb{C}^{N}\right)  }\text{.}%
\]
Then
\begin{align*}
\rho_{\vec{\ell}\vec{m}}  & =\sum_{\vec{j}\in\mathbb{Z}_{N_{1}}^{2}%
}\mathcal{F}_{\vec{j}\vec{\ell}}\mathcal{\bar{F}}_{\vec{j}\vec{m}}\\
& =\frac{1}{N}\sum_{j_{1}=0}^{N_{1}-1}\sum_{j_{2}=0}^{N_{1}-1}\left(
\begin{array}
[c]{l}%
\exp\left(  \frac{2\pi i}{N}\left(  N_{2}j_{1}+\ell_{1}\right)  \left(
N_{2}j_{2}+\ell_{2}\right)  \right) \\
\times\exp\left(  -\frac{2\pi i}{N}\left(  N_{2}j_{1}+m_{1}\right)  \left(
N_{2}j_{2}+m_{2}\right)  \right)
\end{array}
\right) \\
& =\frac{1}{N}\exp\left(  \frac{2\pi i}{N}\left(  \ell_{1}\ell_{2}-m_{1}%
m_{2}\right)  \right) \\
& \times\sum_{j_{1}=0}^{N_{1}-1}\exp\left(  \frac{2\pi i}{N_{1}}\left(
\ell_{2}-m_{2}\right)  j_{1}\right)  \times\sum_{j_{2}=0}^{N_{1}-1}\exp\left(
\frac{2\pi i}{N_{1}}\left(  \ell_{1}-m_{1}\right)  j_{2}\right)
\end{align*}
Evaluating the appropriate inverse-Fourier transforms,
\begin{equation}
\rho_{\vec{\ell}\vec{m}}\equiv\frac{N_{1}}{N_{2}}\exp\left(  \frac{2\pi i}%
{N}\left(  \ell_{1}\ell_{2}-m_{1}m_{2}\right)  \right)  \times\chi
_{N_{1}\mathbb{Z}^{2}}\left(  \vec{\ell}-\vec{m}\right)
\text{.\label{fredden}}%
\end{equation}
The spectral decomposition of $\rho$ into a linear combination of projections
may be simply read off from the asymptotic $n\rightarrow\infty$ behavior of
$\left(  \ref{fredden}\right)  $ to the power of $n\in\mathbb{Z}^{+}%
$.\footnote{using matrix multiplication} One need not do this, however, for
using the identity
\[
\chi_{N_{1}\mathbb{Z}^{2}}\left(  \vec{\ell}-\vec{m}\right)  =\sum
_{C\in\mathcal{M}}\chi_{C}\left(  \vec{\ell}\right)  \chi_{C}\left(  \vec
{m}\right)  \text{,}%
\]
equation $\left(  \ref{fredden}\right)  $ may be rewritten as
\[
\rho=\sum_{C\in\mathcal{M}}\frac{N_{1}}{N_{2}}\left|  C\right|  \times\left|
B_{C}\right\rangle \left\langle B_{C}\right|  \text{,}%
\]
where the $B_{C}$ are orthonormal, as noted before. The $A_{C}$ are easily
computed using $\left(  \ref{findas}\right)  $.
\end{proof}

\end{document}